# Integrated diffractive full-Stokes spectro-polarimetric imaging

JINGYUE MA,[1] ZHENMING YU,[1,2,*] ZHENGYANG LI,[1] LIANG LIN,[1] LIMING CHENG,[1] JIAYU DI,[1] TONGSHUO ZHANG,[1] NING ZHAN,[1] AND KUN XU[1,2]

[1]*State Key Laboratory of Information Photonics and Optical Communications, Beijing University of Posts and Telecommunications, Beijing, China, 100876*
[2]*Xiong'an Aerospace Information Research Institute, Xiong'an, China*
*\*yuzhenming@bupt.edu.cn*

**Abstract:** Spectro-polarimetric imaging provides multidimensional optical information acquisition capabilities, offering significant potential for diverse applications. Current spectro-polarimetric imaging systems typically suffer from large physical footprints, high design complexity, elevated costs, or the drawback of requiring replacement of standard components with polarization optics. To address these issues, we propose an integrated diffractive full-Stokes spectro-polarimetric imaging framework that synergistically combines end-to-end designed diffractive polarization spectral element (DPSE) with SPMSA-Net to demonstrate high-performance spectro-polarimetric imaging. The DPSE modulates scene and generates modulated images carrying phase-encoding and polarization information. The modulated images are the input of the SPMSA-NET for the reconstruction of the spectro-polarimetric data cube. The framework achieves an average improvement of 0.78 dB in PSNR and 0.012 in SSIM over existing state-of-the-art algorithms. Based on this framework, our prototype system can simultaneously capture spectral information (400-700 nm) with 10 nm spectral resolution and full-Stokes parameters $\{S_1, S_2, S_3, S_4\}$. Meanwhile, the system provides high spatial resolution of 2252×2252 pixels. Experimental results demonstrate that our system achieves high-fidelity spectral imaging (over 98.9% fidelity) and precise polarization characterization, with a compact architecture (modulation component of merely 2-mm thickness).

## 1. Introduction

Hyperspectral imaging and polarization imaging belong to different domains, capturing distinct dimensional information. Hyperspectral imaging captures narrow-band reflectance signatures that reveal molecular characteristics of materials. In contrast, polarimetric imaging analyzes variations in lightwave polarization states to sensitively detect physical structural properties of material surfaces such as surface roughness, micro-deformations and stress distribution. These two modalities provide complementary scene information from chemical and physical dimensions respectively, demonstrating significant application value in biomedical [1], remote sensing [2] and agriculture [3]. However, conventional optical sensing devices, such as cameras, spectrometers and polarimeters are limited to single-dimensional data acquisition. This limitation constrains their applicability in scenarios requiring high-dimensional information acquisition.

In recent years, advances in fabrication technologies and computational algorithms have significantly accelerated the development of snapshot hyperspectral imaging [4–22]. These breakthroughs enable the integrated spectro-polarimetric imaging systems [23–32]. Spectro-polarimetric imaging systems can be fundamentally categorized into two implementations: joint spectro-polarimetric encoding and separate spectro-polarimetric encoding. Regarding joint spectro-polarimetric encoding, novel-material-based spectro-polarimetric imaging systems [33–38] have attracted significant research attention due to their exceptional capabilities in simultaneous polarization and spectral modulation. However, such systems typically suffer from high production costs. The requirement for simultaneous spectral and polarimetric encoding in such systems also introduces significant design complexity and reconstruction challenges. Furthermore, such systems exhibit limited spatial resolution. In term of separate encoding, the most straightforward implementation employs direct spatial separation of polarization states and spectral channels at the imaging plane for data acquisition [24,26,39]. However, such systems suffer from the inherent resolution-channel trade-off, while their complex optical architectures substantially limit system flexibility and adaptability. An alternative method combines existing HSI systems with polarization modulation devices [40]. This method decouples spectral and polarization modulation, effectively leveraging the inherent high spatial resolution of existing hyperspectral imaging systems. Through high-precision mechanical modulation devices, the systems enable precise polarization-state acquisition. Furthermore, this method substantially reduces the computational challenge of reconstruction algorithms compared to joint-encoding schemes. However, such polarization modulation devices inevitably introduce additional system complexity, thereby hindering its integration.

In this work, we propose an integrated diffractive full-Stokes spectro-polarimetric imaging framework, that combines proposed diffractive polarization spectral element (DPSE) with designed spectro-polarimetric multi-scale attention-Net (SPMSA-Net) to achieve high-performance spectro-polarimetric imaging. In the framework, the DPSE first modulates a scene containing full polarization and spectral information via phase encoding and polarization filtering, generating a spectrally encoded image with a specific polarization state. This encoded image is then fed into the SPMSA-NET to reconstruct the spectro-polarimetric data. Based on the proposed framework, we deploy an integrated full-Stokes spectro-polarimetric imaging system. The proposed system achieves a high spatial resolution of 2252×2252 pixels with 10 nm spectral resolution across the wavelength range of 400-700 nm and full-Stokes parameter $\{S_1, S_2, S_3, S_4\}$ acquisition. Compared with state-of-the-art methods, our method demonstrates high-fidelity spatial and spectral reconstruction, outperforming them with an average improvement of 0.78 dB in PSNR and an improvement of 0.012 in SSIM. The experiments demonstrated that the system could capture high-fidelity hyperspectral images with precise polarization characterization. Specifically, the spectro-polarimetric images captured by the system achieve over 98.9% spectral fidelity and exhibit accurate characterizations of the degree of linear polarization (DoLP), the angle of linear polarization (AoLP), and a strong discrimination capability for circular polarization states. Moreover, the DPSE features a compact 2-mm thickness, offering excellent plug-and-play functionality and enabling low-cost integration into existing imaging systems. We anticipate that this breakthrough will drive spectro-polarimetric imaging toward higher quality and enhanced compatibility, expanding its applicability in a wide range of fields.

## 2. Principle

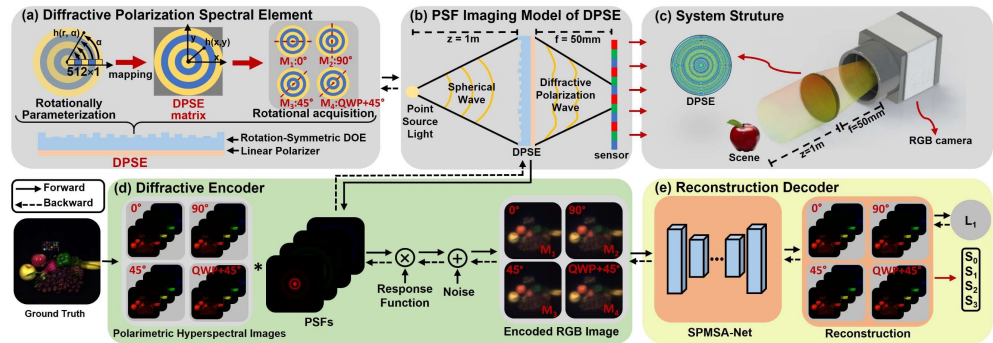

Fig. 1. The integrated diffractive full-Stokes spectro-polarimetric imaging framework. (a) Diffractive polarization spectral element. (b) PSF Imaging model of DPSE. (c) The system structure. (d) Diffractive encoder. (e) Reconstruction decoder.

The integrated diffractive full-Stokes spectro-polarimetric imaging framework is shown in Fig. 1. The framework proposed an innovative spectro-polarimetric encoding component termed as diffractive polarization spectral element (DPSE), illustrated in Fig. 1(a). The DPSE, serving as the diffractive spectro-polarimetric modulation component, generates the point spread functions (PSFs) within the model illustrated in Fig. 1(b). The system structure in Fig. 1(c) is constructed based on the model presented in Fig. 1(b). Fig. 1(d) illustrates the diffractive encoder of the system, which utilizes the PSFs generated in Fig. 1(b) to encode hyperspectral images. The encoded image generated by the encoder is fed into the reconstruction decoder shown in Fig. 1(e), recovering a series of hyperspectral images with polarization information through proposed SPMSA-Net.

## 2.1 Diffractive polarization spectral element

Fig. 1(a) illustrates the composition of the DPSE. DPSE consists of a stepped rotation-symmetric diffractive optics element (DOE) [41] and a linear polarizer. The DOE comprises a lithographic patterned layer with a specific height distribution and a planar non-lithographic layer. The linear polarizer is cemented to the surface of its non-lithographic layer. The DOE performs phase-encoding of incident light across all polarization states. Subsequently, the linear polarizer filters the light aligned with its transmission axis. The height distribution, as shown in Fig.1(a), is generated by rotating a trainable 512 × 1 float height parameter $w_i$, where $i \leq 512, i \in \mathbb{N}$. Specifically, for a 1024 × 1024 height distribution matrix of the DPSE, the center point of the matrix is defined as the origin of the polar coordinate system, with the zero-degree reference axis aligned parallel to one side of the matrix. The height map of DPSE can be represented as:

$$h(r,\alpha) = \begin{cases} w_{round(r)}, & 0 \leq r \leq 512 \\ 0, & r > 512 \end{cases} \quad (1)$$

where $r$ represents the Euclidean distance from any point in the matrix to the origin and $0 \leq \alpha < 2\pi$. This distribution results in an annular height map, endowing the DPSE with rotational symmetry. For the convenience of subsequent calculations, the height map $h(r,\alpha)$ defined in polar coordinates is projected onto a Cartesian coordinate system shown in Fig. 1(a), denoted as $h(x,y)$. This projection preserves the origin while aligning the zero-degree reference axis with the positive x-axis and assigning the perpendicular direction to the positive y-axis.

During acquisition process, we define the horizontal plane of the experimental bench as the reference (0º). As shown in Fig. 1(a), for each scene, we rotate the DPSE and acquire polarization measurements at three polarization angles: 0º, 90º, and 45º, respectively denoted as $M_1$, $M_2$, and $M_3$. Then, we fix the DPSE polarization angle at 45º and place a quarter-wave plate (QWP) in front of the DPSE, with its fast axis aligned to 0º, recording the measurement as $M_4$. Consequently, we obtain four RGB measurements, denoted as $\{M_1, M_2, M_3, M_4\}, M_i \in \mathbb{R}^{H \times W \times 3}$, where $H$ is the height of the image and $W$ is the width of the image. Due to the rotational symmetry of the height distribution, its PSFs remain unchanged during the rotation of the DPSE itself. Moreover, since the polarizer in the DPSE only affects the polarization state of the transmitted light without altering the phase-encoding generated by the front-end DOE, a unified PSF can be employed as the encoder to generate the encoded image. This also enables all encoded images to be reconstructed through a single spectral reconstruction network. Using the end-to-end jointly trained SPMSA-Net, we directly reconstruct the measurements to obtain the corresponding results $\{P_1, P_2, P_3, P_4\}$, $P_i \in \mathbb{R}^{H \times W \times C_\lambda}$, where $C_\lambda = 31$ is the spectral channels. $P_i, i = 1, 2, 3, 4$ contain polarization information. Then, the Stokes parameters can be reconstructed according to:

$$\begin{cases} S_0 = P_1 + P_2 \\ S_1 = P_1 - P_2 \\ S_2 = 2P_3 - S_0 \\ S_3 = S_0 - 2P_4 \end{cases} \quad (2)$$

*2.2 Diffractive spectro-polarimetric imaging model*

The diffractive spectro-polarimetric imaging model comprises two integral components: the PSF imaging model of the DPSE and the diffractive encoder. Fig. 1(b) illustrates the PSF imaging model of the DPSE. The point source light of wavelength $\lambda$ from $z$ can be modeled as a spherical wave:

$$U_0(x, y, \lambda, z) = \exp\left[ik\frac{(x^2 + y^2)}{z}\right] \quad (3)$$

where $(x, y)$ represents the spatial coordinate on the diffractive encoding element, $k = 2\pi / \lambda$ represents the wave number. Then, the wave passes through the DPSE plane, and is changed to:

$$U_1(x, y, \lambda, z) = A(x, y) \cdot \exp\left[ik\left(\frac{(x^2 + y^2)}{z} + (n_\lambda - 1)h(x, y)\right)\right] \quad (4)$$

where $A(x, y)$ is the amplitude aperture function of DPSE, which is treated as a constant $A_0$, $n_\lambda$ is the refractive index of the silicon dioxide layer for wavelength $\lambda$. The wave then passes through the linear polarizer of DPSE, transmitting only the polarization component aligned with the polarizer's transmission axis, denoted as $U_2(x, y, \lambda, z)$. Next, the wave propagates to the sensor by the focal length $f$, resulting in the propagating phase $(x^2 + y^2)/2f$. The wave is then is changed to:

$$U_3(x, y, \lambda, z) = A(x, y) \cdot \exp\left[ik\left(\frac{(x^2 + y^2)}{z} + (n_\lambda - 1)h(x, y) + \frac{(x^2 + y^2)}{2f}\right)\right] \quad (5)$$

The PSFs of the system is then modeled as:

$$P_{psf}(x, y, \lambda, z) = \left|\mathcal{F}\{U_3(x, y, \lambda, z)\}\right|^2 \quad (6)$$

where $\mathcal{F}$ is the Fourier Transform.

Fig. 1(d) shows the diffractive encoder of the framework. In the diffractive encoder, the PSFs of the system is convolved with the hyperspectral image $I(x, y, \lambda)$. We set the $z = 1\text{m}$ and $f = 50$ mm to simplify the calculation. The resulting data is then mapped through the response function $R(\lambda) = T_{polarizer}(\lambda) \cdot R_c(\lambda)$, where $T_{polarizer}(\lambda)$ represents the transmittance function of polarization components and $R_c(\lambda)$ represents the response function of RGB camera. The encoded image $I_{c \in \{R,G,B\}}(x, y)$ is modeled as:

$$I_{c \in \{R,G,B\}}(x, y) = \sum_{\lambda = \lambda_0}^{\lambda_i} (P(x, y, \lambda, z_0) \otimes I(x, y, \lambda))R(\lambda) + n \quad (7)$$

where $\lambda_0$ is the minimum wavelength, $\lambda_i$ is the maximum wavelength, and $n$ is the sensor noise.

## 2.3 The reconstruction decoder with SPMSA-Net

Fig. 1(e) illustrates the reconstruction decoder of the framework. We design a reconstruction network, termed as the SPMSA-Net, to reconstruct the hyperspectral data from the encoded image of the system. Fig. 2 illustrates the overall architecture of the SPMSA-Net along with the detailed structures of its individual modules.

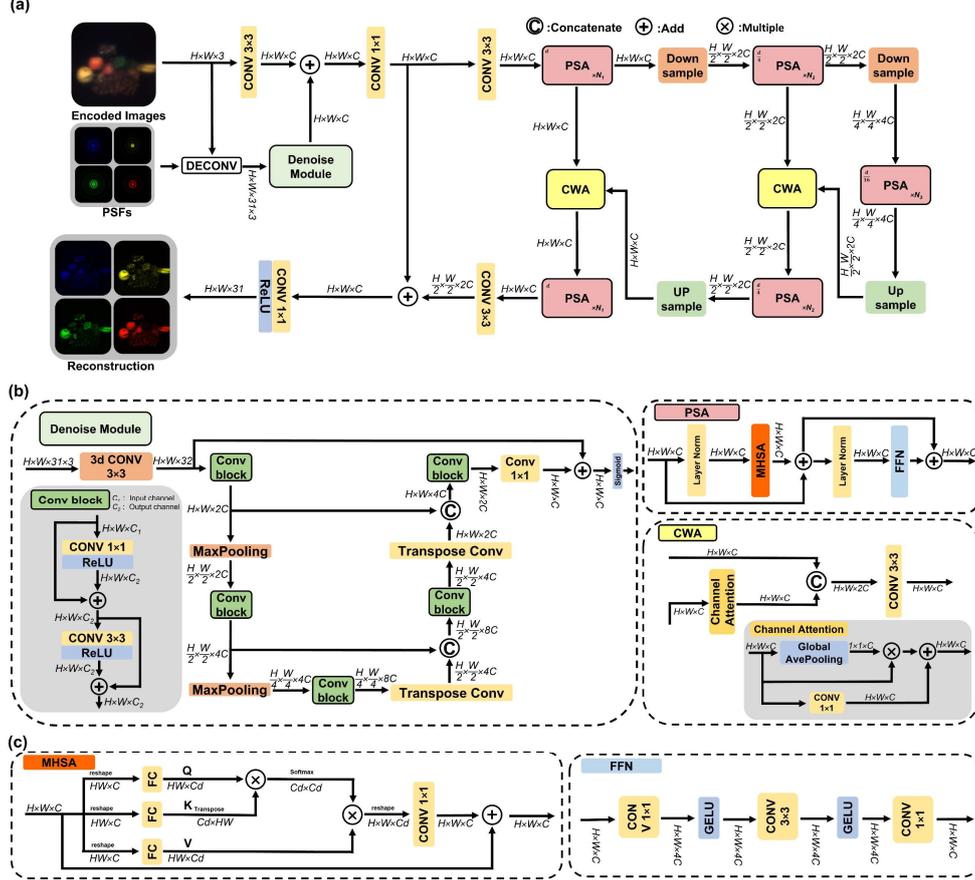

Fig. 2. (a) The framework of the SPMSA-Net. (b) Denoise module, Polarization-Spectral Attention module and Channel attention module. (c) Multi-Head Self-Attention module and Feed-Forward Network module

Fig. 2(a) presents the overall architecture of the SPMSA-Net. In Fig. 2(a), both the encoded images and PSFs are fed into the network. The encoded images first undergo the 'DECONV' module, where it is jointly processed with the PSFs through deconvolution algorithm. Since the encoded images are generated according to Eq. (7), the hyperspectral image of each wavelength band can be reconstructed by deconvolving the encoded image with the corresponding PSF, as expressed mathematically as:

$$\hat{I}_{\{R,G,B\}}(\lambda) = deconv(I_{c\in\{R,G,B\}}, PSF_{\lambda,c\in\{R,G,B\}}) = F^{-1}\left(\frac{F(I_{c\in\{R,G,B\}}) \cdot F(P(x,y,\lambda,z_0))^*}{|F(P(x,y,\lambda,z_0))|^2}\right) \quad (8)$$

For a specific wavelength band $\lambda_j$, the RGB map of the hyperspectral image $I_{\{R,G,B\}}(x,y,\lambda_j)$ can be expressed as the deconvolution of the encoded image with the PSF at wavelength $\lambda_j$, subtracted by the deconvolution results of this PSF with measurements at other wavelengths:

$$I_{\{R,G,B\}}(x,y,\lambda_j) = \hat{I}_{\{R,G,B\}}(\lambda_j) - \sum_{\lambda=\lambda_0,\lambda \neq \lambda_j}^{\lambda_i} F^{-1}(\frac{F(I_{c\in\{R,G,B\}}(x,y,\lambda)) \cdot F(P(x,y,\lambda_j,z_0))^*}{|F(P(x,y,\lambda_j,z_0))|^2}) \quad (9)$$

By treating the deconvolution results from mismatched wavelength bands in Eq. (9) as noise, the expression can be simplified to:

$$I_{\{R,G,B\}}(x,y,\lambda_j) = \hat{I}_{\{R,G,B\}}(\lambda_j) - n_{deconvolution} \quad (10)$$

After processing each wavelength band, the 'DECONV' module outputs an $H \times W \times 31 \times 3$ tensor with noise $n_{deconvolution}$ on each wavelength. The resulting tensor is then processed by the 'Denoising Module' to generate a $H \times W \times C$ feature map ($C = 32$), which is subsequently summed with another $H \times W \times C$ feature map derived from the encoded image via a 3×3 convolution layer. The summed feature map is then processed by a 1×1 convolutional layer to reweight channels adaptively. The output of this stage is denoted as $O_1$, which serves as the input to the subsequent network. Inspired by MST++ [42], we developed a U-shaped Transformer architecture to extract deep features from $O_1$. First, $O_1$ is processed by a 3×3 convolution as embedding layer and processed through $N_1 = 2$ Polarization-Spectral Attention (PSA) modules, denoted as $O_2$. Then, $O_2$ pass through a 4×4 strided convolutional layer as downsampling, and subsequently through $N_2 = 2$ additional PSA modules, ultimately generating the output feature map $O_3$. $O_3$ undergoes downsampling followed by processing through $N_3 = 2$ PSA modules, and is subsequently upsampled via a strided transposed convolution layer with 2×2 kernels, yielding the output feature map $O_4$. Next, $O_3$ and $O_4$ are jointly fed into a 'Channel Weight Attention (CWA) Module' to integrate features extracted from different scales. The output of the CWA module is subsequently processed through $N_2$ PSA modules and upsampled via a 2×2 strided transposed convolution layer, yielding the output feature map $O_5$. $O_2$ and $O_5$ are jointly sent to a CWA module followed by $N_1$ PSA module and a mapping 3×3 convolutional layer, denoted as $O_6$. Finally, the reconstruction is obtained by summing feature maps $O_1$ and $O_6$, followed by a 1×1 convolutional layer and ReLU activation.

The detailed architecture of the 'Denoising Module' is illustrated on the left part Fig. 2(b). The 'Denoising Module', implemented as a U-Net architecture [43], is used to suppresses noise components $n_{deconvolution}$ while preserving critical features for downstream network processing. The denoising module employs designed residual convolution blocks [44], termed as 'conv block', with down sampling and up sampling operations implemented through max-pooling and transposed convolution layers respectively, enabling multi-scale feature extraction. The details of 'PSA module' are shown in the top right part of the Fig. 2(b). We deploy Layer Normalization, 'Multi-Head Self-Attention (MHSA) Module', 'Feed-Forward Network (FFN) Module', and residual connections in 'PSA module'.

The details of designed CWA module are shown in the bottom right part of the Fig. 2(b). We design a channel attention module to adaptively reweight feature maps from deep layers. The channel attention module first applies global average pooling to transform the $H \times W \times C$ input features into $1 \times 1 \times C$ channel-wise weights. The original $H \times W \times C$ feature map is then multiplied by these weights. In parallel, the input features undergo a 1×1 convolution layer, and the resulting feature maps are combined with the reweighted feature map via residual addition. This architecture enables simultaneous global channel weighting and local feature preservation.

Fig. 2(c) illustrates the architecture of the MHSA and FFN modules. The MHSA module adopts the multi-head attention computation augmented with a 1×1 convolutional layer and residual connections after the attention mechanism to enhance feature extraction. The FFN module maintains an identical structure to MST++ [42].

The loss function of the network is L1 loss (Mean Absolute Error, MAE) between the reconstruction hyperspectral images and the ground truth. The network is trained for a total of 150 epochs. The learning rate of the network is $lr_{network} = 0.0005$, which is multiplied by 0.9 every 30 epochs. The training dataset is CAVE [45] dataset.

## 3. Experimental setup

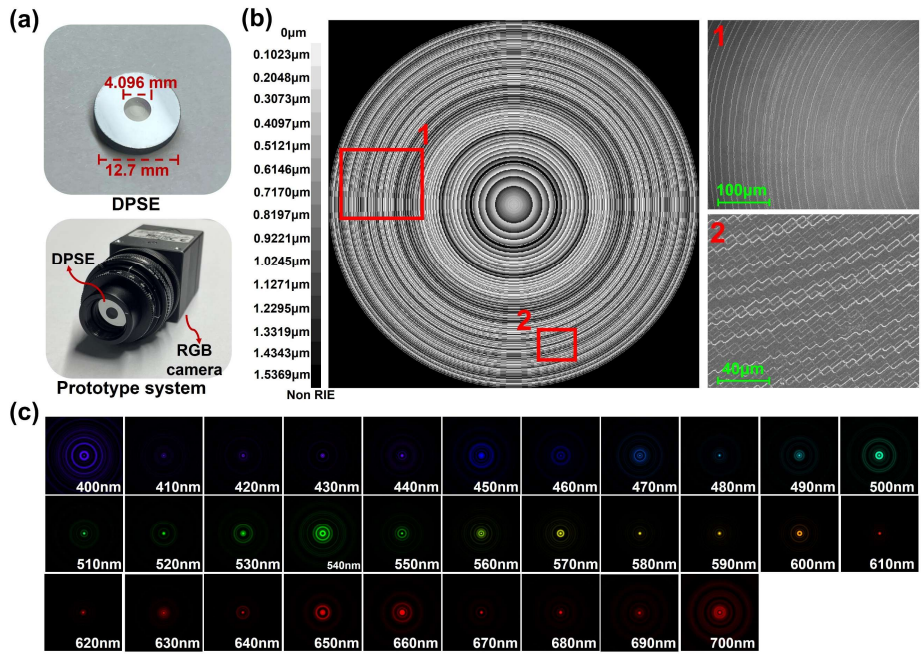

Fig. 3. (a) The DPSE and the prototype system. (b) The height map and SEM images of the DPSE. (c) The PSFs of the DPSE.

The DPSE and the experimental prototype system are shown in Fig. 3(a). The DPSE has a 12.7 mm outer diameter, featuring 4.096 mm diffractive apertures surrounded by chromium-coated light-blocking regions to minimize stray light interference. The prototype system consists of a DPSE coupled with a RGB camera (FLIR BFS-U3-200S7C-C). The DPSE is mounted on a continuous rotation mount that enables continuous 360º angular positioning, while maintaining a fixed 50.0 mm working distance between the DPSE and camera sensor through calibrated optomechanical components. This configuration provides both rotational freedom for polarization measurements and precise optical alignment for imaging performance.

The left panel of Fig. 3(b) displays the height map and their distribution, corresponding to the height profile within the diffractive apertures shown in Fig. 3(a). The DPSE incorporates a 4 μm single-pixel size, with a 1024 × 1024 height map matrix uniformly quantized into 16 steps and a total depth of 1.5369 μm. The DPSE is fabricated using lithography and reactive ion etching (RIE) techniques. The manufacturing error is 40 nm for each step. The right panel of Fig. 3(b) presents scanning electron microscopy (SEM) images of two selected regions in the height map (Region 1 and Region 2). Fig. 3(c) shows the PSFs of the DPSE across the wavelength range of 400-700 nm. These spectral bands exhibit a one-to-one correspondence with the bands of hyperspectral imaging results. It is obvious that the height map, SEM images and PSFs of DPSE collectively demonstrate pronounced rotational symmetry.

During the experiment, a halogen lamp is used to illuminate the scene. Since the DPSE's polarization-selective transmission inherently attenuates at least 50% of incident light intensity, we perform 2×2 pixels binning on the original 4504×4504 sensor data, resulting in a final resolution of 2252×2252 pixels to maintain sufficient signal-to-noise ratio. The captured images were in 12-bit RAW format. The single-frame exposure time is set to 40 ms, and a total of 4 frames are acquired for each scene.

## 4. Results

### 4.1 Simulation analysis

To evaluate the spectral reconstruction capability of the SPMSA-Net, ten distinct scenes are selected from each of the ICVL [46] and KAUST [47] datasets for simulation analysis. We downsample the ICVL dataset to a spatial resolution of 512×512 pixels for evaluation. Fig. 4 presents these the RGB reference of selected images from two datasets.

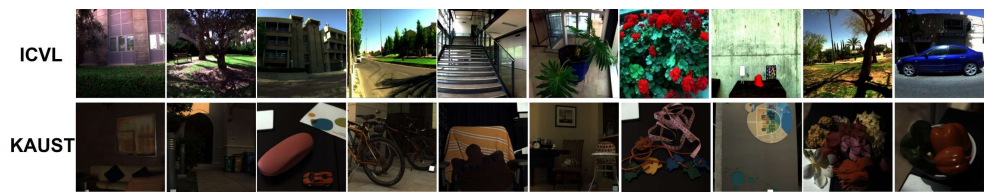

Fig.4 RGB reference of selected images from two datasets.

Table 1 and Table 2 lists the PSNR and SSIM of scenes in ICVL and KAUST datasets by using Res-UNet [41], HD-Net [48], MST++ [42], CAFormer [12] and proposed SPMSA-Net. PSNR and SSIM values in bold represent the best results for each scene. It is obvious that our proposed SPMSA-Net outperforms other algorithms across most scenarios. Compared with state-of-the-art methods, our approach achieves a performance gain of 1.22 dB in PSNR and 0.01 in SSIM on the ICVL dataset, and 0.34 dB in PSNR and 0.014 in SSIM on the KAUST dataset. Totally, the system demonstrates an average gain of 0.78 dB in PSNR and an improvement of 0.012 in SSIM.

Table 1. PSNR (dB) and SSIM of 10 ICVL test scenes reconstructed by different methods

|         | Res-Unet      | HDNet         | MST++         | CAFormer      | SPMSA-Net       |
|---------|---------------|---------------|---------------|---------------|-----------------|
| Scene 1 | 28.67 / 0.930 | 28.73 / 0.933 | 29.25 / 0.936 | 30.82 / 0.969 | **32.59 / 0.979** |
| Scene 2 | 31.67 / 0.862 | 31.50 / 0.860 | 32.06 / 0.870 | 34.03 / 0.937 | **35.78 / 0.962** |
| Scene 3 | 32.68 / 0.936 | 32.73 / 0.939 | 33.15 / 0.944 | 34.17 / 0.969 | **34.78 / 0.977** |
| Scene 4 | 34.49 / 0.934 | 34.96 / 0.940 | 35.39 / 0.942 | 35.91 / 0.965 | **37.48 / 0.977** |
| Scene 5 | 33.56 / 0.941 | 33.57 / 0.945 | 33.77 / 0.947 | 36.10 / 0.975 | **38.00 / 0.982** |
| Scene 6 | 33.46 / 0.941 | 33.57 / 0.944 | 34.33 / 0.951 | 35.05 / 0.973 | **36.09 / 0.977** |
| Scene 7 | 32.10 / 0.950 | 32.76 / 0.955 | 33.04 / 0.958 | 33.33 / 0.973 | **34.92 / 0.979** |
| Scene 8 | 31.67 / 0.897 | 31.15 / 0.895 | 31.74 / 0.905 | **33.38 / 0.947** | 33.14 / 0.946 |
| Scene 9 | 32.96 / 0.884 | 33.54 / 0.890 | 33.84 / 0.893 | 34.70 / 0.938 | **35.55 / 0.959** |
| Scene 10| 33.78 / 0.931 | 34.59 / 0.939 | 34.65 / 0.940 | 35.45 / 0.960 | **36.78 / 0.973** |
| Average | 32.50 / 0.921 | 32.71 / 0.924 | 33.12 / 0.929 | 34.29 / 0.961 | **35.51 / 0.971** |

Table 2. PSNR (dB) and SSIM of 10 KAUST test scenes reconstructed by different methods

|         | Res-Unet      | HDNet             | MST++          | CAFormer        | SPMSA-Net         |
|---------|---------------|-------------------|----------------|-----------------|-------------------|
| Scene 1 | 35.77 / 0.877 | **38.22** / 0.910 | 38.06 / 0.911  | 36.47 / 0.916   | 37.04 / **0.931** |
| Scene 2 | 31.11 / 0.888 | **36.21** / 0.898 | 34.97 / 0.894  | 32.11 / 0.939   | 31.92 / **0.957** |
| Scene 3 | 30.49 / 0.882 | 31.31 / 0.886     | 31.59 / 0.892  | 31.62 / 0.910   | **33.36 / 0.948** |
| Scene 4 | 29.50 / 0.940 | 31.99 / 0.951     | 31.81 / 0.958  | 31.09 / **0.967** | 30.69 / 0.962   |
| Scene 5 | 30.85 / 0.895 | 30.84 / 0.904     | 31.33 / 0.913  | 31.68 / **0.938** | **32.69** / 0.937 |
| Scene 6 | 38.78 / 0.931 | 38.86 / 0.939     | 38.67 / 0.940  | 39.21 / 0.959   | **39.83 / 0.967** |
| Scene 7 | 31.47 / 0.898 | 31.75 / 0.904     | 32.19 / 0.909  | 32.59 / 0.924   | **33.52 / 0.951** |
| Scene 8 | 36.17 / 0.934 | 36.39 / 0.946     | 37.09 / 0.946  | 37.18 / 0.944   | **38.62 / 0.962** |
| Scene 9 | 32.89 / 0.900 | **35.42** / 0.934 | 34.56 / 0.935  | 34.59 / 0.938   | 33.82 / **0.943** |
| Scene 10| 33.38 / 0.875 | 33.95 / 0.880     | 34.47 / 0.889  | 36.07 / 0.945   | **36.79 / 0.961** |
| Average | 33.04 / 0.902 | 34.49 / 0.915     | 34.47 / 0.919  | 34.26 / 0.938   | **34.83 / 0.952** |

Fig. 5 presents the spectral curves and visual comparisons of different algorithms at a resolution of 512×512 pixels. Fig. 5(a) selects two points (P1 and P2) from scenes in the ICVL and KAUST datasets respectively and plots their corresponding spectral curves. The spectral curve at point P1 demonstrates that our algorithm achieves high-fidelity reconstruction of the scene, particularly capturing the spectral peak accurately within the 640–670 nm wavelength range. For point P2, it can be observed that our reconstruction aligns closely with the ground truth in wavelengths below 500 nm, while exhibiting a smoother spectral response compared to other algorithms. Fig. 5(b) displays the spatial reconstruction results of a selected scene across multiple spectral bands (450 nm, 510 nm, 590 nm, 640 nm). It can be observed that within the zoomed regions, our proposed SPMSA-Net delivers sharper reconstructions than alternative approaches.

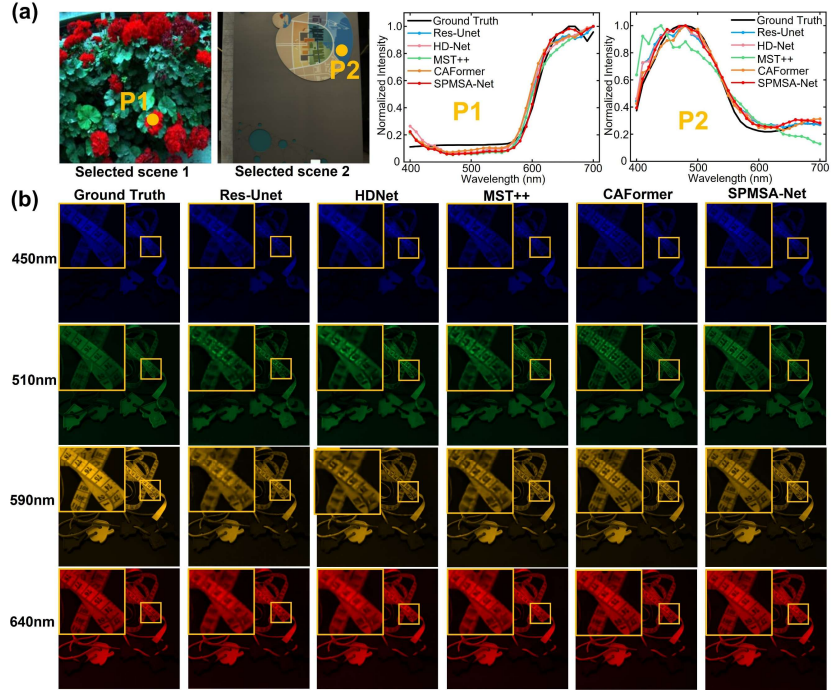

Fig. 5 The simulation results of different algorithms. (a) The selected scenes and the spectral curve of two points. (b) The spatial reconstruction results of different algorithms.

*4.2 Experimental results*

To validate the experimental spectral imaging capability of the system, we acquired images of a standard color checker with the DPSE fixed at 0º polarization orientation. Fig. 6 illustrates the spectral imaging results of the system. The left part of Fig. 6(a) shows the 'Measurement' and the 'Synthesized RGB' images. The 'Measurement' image is the original encoded measurement of the system. The acquired measurement is subsequently fed into the SPMSA-Net for reconstruction to obtain the hyperspectral data. The 'Synthesized RGB' image is the synthesized RGB image of reconstruction hyperspectral data. The right part of Fig. 6(a) shows spectral curves from two selected points (P1: green, P2: red). We compared our algorithm with the existing state-of-the-art algorithms [12,41,42,48]. In the spectral curve of Point P1, SPMSA-Net accurately reconstructed the reflectance peak of the green patch within the 500-600 nm wavelength range, outperforming other comparative methods significantly. Regarding Point P2, the spectral curve demonstrates that the reconstruction achieved by SPMSA-Net is not only smoother overall but also aligns more closely with the reference data, highlighting its exceptional spectral reconstruction capability. For P1 and P2, we calculated the Mean Squared Error (MSE) between the spectral curves reconstructed by all algorithms and the ground truth. It can be seen that our method achieves the lowest MSE, demonstrating its superior performance in spectral reconstruction. We evaluate the fidelity of the DPSE reconstruction against the ground truth data through the spectral cosine similarity [49]. The fidelity values are 98.93% for P1 and 99.70% for P2. This demonstrates that our proposed method achieved high spectral fidelity in the scene. Fig. 6(b) presents spectral imaging results across 400-700nm wavelengths, confirming excellent spatial quality in reconstructed spectral channels.

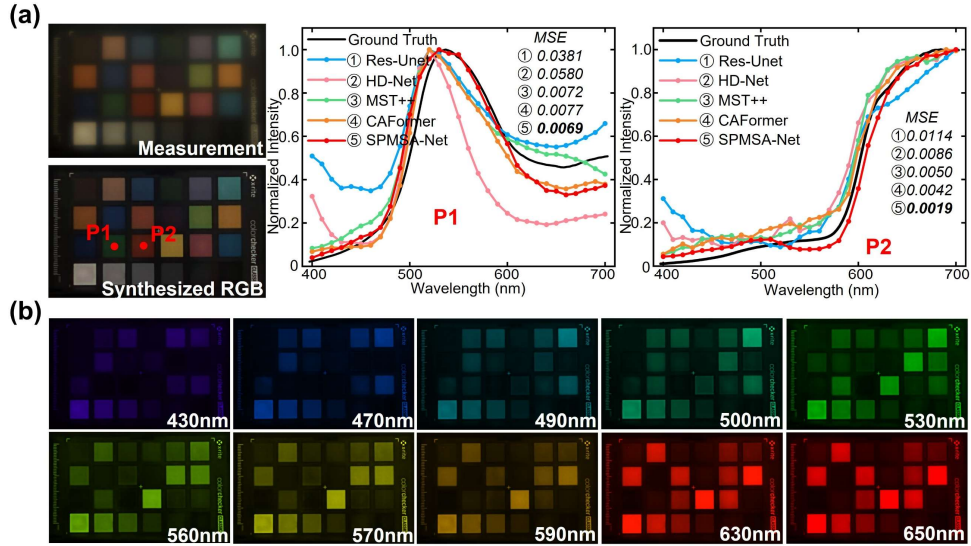

Fig. 6 The hyperspectral imaging results. (a) The measurement, the synthesized RGB images and the spectral curve of the scene. (b) The spectral imaging results across 400-700 nm.

We deployed a series of polarization sensitive scenes to validate the full-Stokes spectro-polarimetric imaging capability of the system. Fig. 7 illustrates the full-Stokes spectro-polarimetric imaging results. We first design a linear polarization test target featuring the 'BUPT' logo with each colored letter covered by a rectangular linear polarizer. The transmission axes of these polarizers are aligned parallel to their long edges, which are fixed at 0º (letter 'T'), 45º (letter 'U'), 90º (letter 'P') and -45º (letter 'B') respectively. Fig. 7(a) shows the synthesized RGB images of different acquisition angles (0º, 45º, 90º, QWP+45º) from reconstruction data. As the DPSE rotates, the linear polarizer in the synthesized RGB also exhibits corresponding changes in brightness. Fig. 7(b) shows the $S_0$ image of different wavelengths that calculated by Eq. (2). The red letter 'B' and orange letter 'U' gradually brighten with increasing wavelength, while the green letter P first brightens and then darkens, and the blue letter T progressively darkens, which aligns with the characteristic spectral bands corresponding to each respective color. Fig. 7(c) shows the $S_1$, $S_2$ images, the DoLP and AoLP at 590nm. The measured image with Stokes parameters exhibits the expected polarization signatures: horizontally and vertically oriented polarizers in the test target produce positive and negative $S_1$ values respectively, while those oriented at +45º and -45º present positive and negative $S_2$ values. The DoLP is calculated using $DoLP = \sqrt{S_1^2 + S_2^2}/S_0$, while the AoLP is calculated using $AoLP = (1/2)\arctan(S_2/S_1)$. In the DoLP image, the polarizer regions exhibit high brightness values approaching +1, clearly indicating the system's effective discrimination of linearly polarized light. The AoLP image shows the linear polarization angles of the scene. Since light reflected or scattered from scene surfaces often carries a degree of polarization state, the AoLP exhibits significant fluctuations. In the AoLP image, the polarization regions exhibit significant polarization uniformity and align closely with their intended orientation. To demonstrate the accuracy of the AoLP, two points (Point 1 and Point 2) are selected at the polarizer positions within the AoLP image. The measured AoLP values are 0.2453 $\pi$ (44.15º) for Point 1 and 0.0071 $\pi$ (1.27º) for Point 2. It is obviously that the AoLP values of two points are all close to the ground truth in the scenario. The results of both the DoLP and the AoLP demonstrate that the system is capable of acquiring linear spectro-polarimetric information with high accuracy.

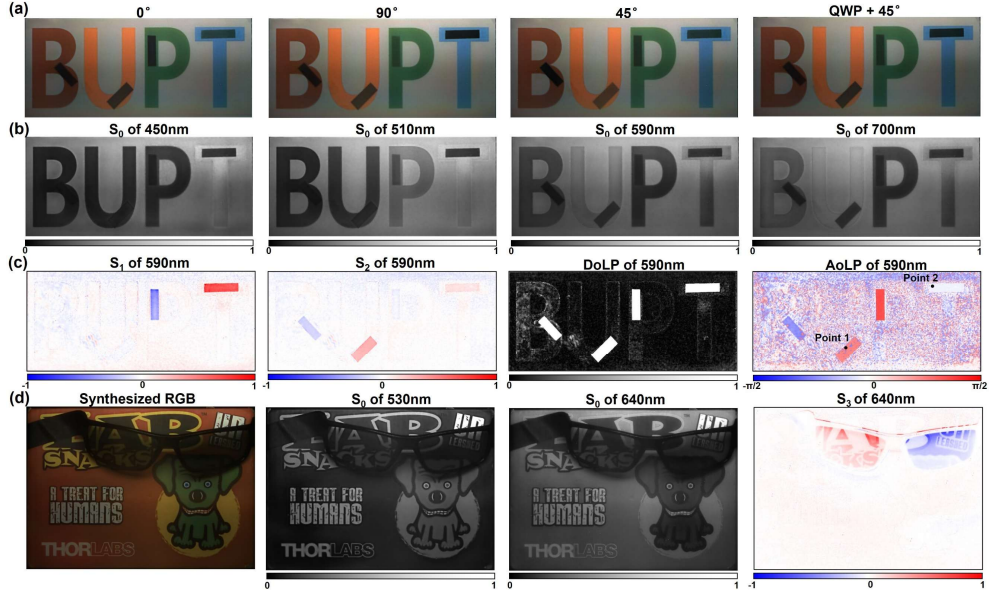

Fig. 7 The full-Stokes spectro-polarimetric imaging results. (a) The synthesized RGB images of different acquisition angles for linear polarization scene. (b) The $S_0$ parameter of different wavelengths. (c) The $S_1$, $S_2$ parameter and the DoLP and AoLP of the 590nm. (d) The synthesized RGB images of circular polarization scene and the $S_0$, $S_3$ parameter.

Fig. 7(d) shows the acquisition results from another experimental scene consisting of a Thorlabs box and a pair of 3D glasses. The left and right lenses of the glasses filter right-circularly polarized (RCP) and left-circularly polarized (LCP) light respectively. The synthesized RGB image is shown in Fig. 7(d) with high spatial resolution. $S_0$ image at 530 nm and 640 nm wavelengths are shown to demonstrate the spectral imaging capability: green patches exhibit dominant response at 530 nm, while red patches show dominant response at 640 nm. $S_3$ images at 640 nm show distinct circular polarization transmission: The left lens predominantly transmits RCP light, while the right lens exhibits LCP transmission. This is consistent with the designed optical properties of the 3D glasses and demonstrates the capability of sytem to discriminate circular polarization states.

## 5. Conclusion

In this work, we propose an integrated diffractive spectro-polarimetric imaging framework which achieves joint optimization by designed DPSE and SPMSA-Net. Compared to existing state-of-the-art algorithms, our framework achieves significant gains, with an average improvement of 0.78 dB in PSNR and 0.012 in SSIM. Based on this framework, we demonstrated a high-performance spectro-polarimetric imaging system that operates across the visible spectrum (400–700 nm) with 10 nm spectral resolution and maintains a high spatial resolution of 2252×2252 pixels. Crucially, our framework enables full-Stokes spectro-polarimetric imaging, accurately reconstructing spectral data over 98.9% fidelity and exhibiting $\{S_0, S_1, S_2, S_3\}$ parameters of each spectral band while also characterizing DoLP and AoLP with high precision. In contrast to the large footprint of scanning systems, the design complexity and high cost of joint spectro-polarimetric encoding systems, and the inadequacy of snapshot spectral imaging systems which require polarization component swapping for polarimetric acquisition, our proposed DPSE integrates both spectral and polarization encoding functions into a single component merely 2 mm thick. Experimental validations confirm the system's ability to resolve spectral signatures and polarization states, demonstrating its potential as a powerful tool for applications in remote sensing, biomedical imaging and material characterization.

**Funding.** The Science and Technology Innovation Project for Xiong'an New Area (NO. 2023XAGG0089); National Natural Science Foundation of China (No. 62522502, 62371056); Major Science and Technology Support Program of Hebei Province (No. 252X1701D); Sponsored by Beijing Nova Program；Shenzhen Science and Technology Program (KJZD20230923115202006); the Fund of State Key Laboratory of Information Photonics and Optical Communication BUPT (No. IPOC2025ZZ02); the Fundamental Research Funds for the Central Universities (No. 530424001, 2024ZCJH13)

**Disclosures.** The authors declare no conflicts of interest.

**Data availability.** Data underlying the results presented in this paper are not publicly available at this time but may be obtained from the authors upon reasonable request.

## References

1. G. Lu and B. Fei, "Medical hyperspectral imaging: a review," J. Biomed. Opt **19**, 010901 (2014).
2. R. Näsi, E. Honkavaara, P. Lyytikäinen-Saarenmaa, M. Blomqvist, P. Litkey, T. Hakala, N. Viljanen, T. Kantola, T. Tanhuanpää, and M. Holopainen, "Using UAV-Based Photogrammetry and Hyperspectral Imaging for Mapping Bark Beetle Damage at Tree-Level," Remote Sensing **7**, 15467–15493 (2015).
3. L. M. Dale, A. Thewis, C. Boudry, I. Rotar, P. Dardenne, V. Baeten, and J. A. F. Pierna, "Hyperspectral Imaging Applications in Agriculture and Agro-Food Product Quality and Safety Control: A Review," Applied Spectroscopy Reviews **48**, 142–159 (2013).
4. M. E. Gehm, R. John, D. J. Brady, R. M. Willett, and T. J. Schulz, "Single-shot compressive spectral imaging with a dual-disperser architecture," Optics Express **15**, 14013 (2007).
5. A. Wagadarikar, R. John, R. Willett, and D. Brady, "Single disperser design for coded aperture snapshot spectral imaging," Applied Optics **47**, B44 (2008).
6. D. S. Jeon, S.-H. Baek, S. Yi, Q. Fu, X. Dun, W. Heidrich, and M. H. Kim, "Compact snapshot hyperspectral imaging with diffracted rotation," ACM Trans. Graph. **38**, 1–13 (2019).
7. Z. Yu, D. Liu, L. Cheng, Z. Meng, Z. Zhao, X. Yuan, and K. Xu, "Deep learning enabled reflective coded aperture snapshot spectral imaging," Opt. Express **30**, 46822 (2022).

8. J. Xiong, X. Cai, K. Cui, Y. Huang, J. Yang, H. Zhu, W. Li, B. Hong, S. Rao, Z. Zheng, S. Xu, Y. He, F. Liu, X. Feng, and W. Zhang, "Dynamic brain spectrum acquired by a real-time ultraspectral imaging chip with reconfigurable metasurfaces," Optica **9**, 461 (2022).
9. J. Yang, K. Cui, X. Cai, J. Xiong, H. Zhu, S. Rao, S. Xu, Y. Huang, F. Liu, X. Feng, and W. Zhang, "Ultraspectral Imaging Based on Metasurfaces with Freeform Shaped Meta-Atoms," Laser & Photonics Reviews **16**, (2022).
10. M. Yako, Y. Yamaoka, T. Kiyohara, C. Hosokawa, A. Noda, K. Tack, N. Spooren, T. Hirasawa, and A. Ishikawa, "Video-rate hyperspectral camera based on a CMOS-compatible random array of Fabry–Pérot filters," Nat. Photon. **17**, 218–223 (2023).
11. W. Zhang, J. Suo, K. Dong, L. Li, X. Yuan, C. Pei, and Q. Dai, "Handheld snapshot multi-spectral camera at tens-of-megapixel resolution," Nat Commun **14**, 5043 (2023).
12. N. Xu, H. Xu, S. Chen, H. Hu, Z. Xu, H. Feng, Q. Li, T. Jiang, and Y. Chen, "Snapshot hyperspectral imaging based on equalization designed DOE," Opt. Express **31**, 20489 (2023).
13. J. Yang, K. Cui, Y. Huang, W. Zhang, X. Feng, and F. Liu, "Deep-learning based on-chip rapid spectral imaging with high spatial resolution," Chip **2**, 100045 (2023).
14. Z. Shi, I. Chugunov, M. Bijelic, G. Côté, J. Yeom, Q. Fu, H. Amata, W. Heidrich, and F. Heide, "Split-Aperture 2-in-1 Computational Cameras," ACM Trans. Graph. **43**, 1–19 (2024).
15. Z. Shi, X. Dun, H. Wei, S. Dong, Z. Wang, X. Cheng, F. Heide, and Y. Peng, "Learned Multi-aperture Color-coded Optics for Snapshot Hyperspectral Imaging," ACM Trans. Graph. **43**, 1–11 (2024).
16. L. Bian, Z. Wang, Y. Zhang, L. Li, Y. Zhang, C. Yang, W. Fang, J. Zhao, C. Zhu, Q. Meng, X. Peng, and J. Zhang, "A broadband hyperspectral image sensor with high spatio-temporal resolution," Nature **635**, 73–81 (2024).
17. H. Xu, S. Chen, H. Hu, P. Luo, Z. Jin, Q. Li, Z. Xu, H. Feng, Y. Chen, and T. Jiang, "Wavelength encoding spectral imaging based on the combination of deeply learned filters and an RGB camera," Opt. Express **32**, 10741 (2024).
18. H. He, Y. Zhang, Y. Shao, Y. Zhang, G. Geng, J. Li, X. Li, Y. Wang, L. Bian, J. Zhang, and L. Huang, "Meta-Attention Network Based Spectral Reconstruction with Snapshot Near-Infrared Metasurface," Advanced Materials **36**, 2313357 (2024).
19. L. Bian, D. Li, S. Wang, C. Teng, J. Wu, H. Liu, H. Xu, X. Chang, G. Zhao, S. Li, and J. Zhang, "Towards large-scale single-shot millimeter-wave imaging for low-cost security inspection," Nat Commun **15**, 6459 (2024).
20. Y. Fan, W. Huang, F. Zhu, X. Liu, C. Jin, C. Guo, Y. An, Y. Kivshar, C.-W. Qiu, and W. Li, "Dispersion-assisted high-dimensional photodetector," Nature **630**, 77–83 (2024).
21. Z. Zhao, Y. Zhang, J. Han, L. Bai, F. Xiong, J. Lu, S. Gong, X. Ke, and S. Jiang, "Self-supervised constrained super-resolution fast coded spectral imaging system," Opt. Express **33**, 5640 (2025).
22. X. Tan, Y. Zhang, L. Zhang, Y. Zhu, J. Chen, Y. Zhou, G. Liang, Z. Wen, J. Xiang, and G. Chen, "High spatial resolution spectral imaging based on amplitude-phase joint modulation metasurfaces using a global optimization algorithm," Opt. Express **33**, 4765 (2025).
23. T. Sun, J. Hu, X. Zhu, F. Xu, and C. Wang, "Broadband Single-Chip Full Stokes Polarization-Spectral Imaging Based on All-Dielectric Spatial Multiplexing Metalens," Laser & Photonics Reviews **16**, 2100650 (2022).
24. T. Mu, F. Han, H. Li, A. Tuniyazi, Q. Li, H. Gong, W. Wang, and R. Liang, "Snapshot hyperspectral imaging polarimetry with full spectropolarimetric resolution," Optics and Lasers in Engineering **148**, 106767 (2022).
25. Y. Ni, C. Chen, S. Wen, X. Xue, L. Sun, and Y. Yang, "Computational spectropolarimetry with a tunable liquid crystal metasurface," eLight **2**, 23 (2022).
26. F. Han, T. Mu, H. Li, and A. Tuniyazi, "Deep image prior plus sparsity prior: toward single-shot full-Stokes spectropolarimetric imaging with a multiple-order retarder," Adv. Photon. Nexus **2**, (2023).
27. Y. Zhang, H. Li, J. Sun, X. Zhang, and H. Wang, "Spectral Polarization Image Reconstruction Using Compressed Sensing Method," IEEE Trans. Geosci. Remote Sensing **61**, 1–13 (2023).
28. A. Fan, T. Xu, J. Li, G. Teng, X. Wang, Y. Zhang, and C. Xu, "Compressive full-Stokes polarization and flexible hyperspectral imaging with efficient reconstruction," Optics and Lasers in Engineering **160**, 107256 (2023).
29. X. Wang, Z. Yang, F. Bao, T. Sentz, and Z. Jacob, "Spinning metasurface stack for spectro-polarimetric thermal imaging," Optica **11**, 73 (2024).
30. S. Wen, X. Xue, S. Wang, Y. Ni, L. Sun, and Y. Yang, "Metasurface array for single-shot spectroscopic ellipsometry," Light Sci Appl **13**, 88 (2024).
31. X. Wang, T. Van Mechelen, S. Bharadwaj, M. Roknuzzaman, F. Bao, R. Rahman, and Z. Jacob, "Exploiting universal nonlocal dispersion in optically active materials for spectro-polarimetric computational imaging," eLight **4**, 22 (2024).
32. L. Zhang, C. Zhou, B. Liu, Y. Ding, H.-J. Ahn, S. Chang, Y. Duan, M. T. Rahman, T. Xia, X. Chen, Z. Liu, and X. Ni, "Real-time machine learning–enhanced hyperspectro-polarimetric imaging via an encoding metasurface," Sci. Adv. **10**, eadp5192 (2024).
33. K. Shinoda and Y. Ohtera, "Alignment-free filter array: Snapshot multispectral polarization imaging based on a Voronoi-like random photonic crystal filter," Opt. Express **28**, 38867 (2020).
34. T. Sun, J. Hu, X. Zhu, F. Xu, and C. Wang, "Broadband Single-Chip Full Stokes Polarization-Spectral Imaging Based on All-Dielectric Spatial Multiplexing Metalens," Laser & Photonics Reviews **16**, (2022).

35. Z. Lin, R. Pestourie, C. Roques-Carmes, Z. Li, F. Capasso, M. Soljačić, and S. G. Johnson, "End-to-end metasurface inverse design for single-shot multi-channel imaging," Opt. Express **30**, 28358 (2022).
36. Y. Fan, W. Huang, F. Zhu, X. Liu, C. Jin, C. Guo, Y. An, Y. Kivshar, C.-W. Qiu, and W. Li, "Dispersion-assisted high-dimensional photodetector," Nature **630**, 77–83 (2024).
37. Q. Zhang, P. Lin, C. Wang, Y. Zhang, Z. Yu, X. Liu, Y. Lu, T. Xu, and Z. Zheng, "Neural-Optic Co-Designed Polarization-Multiplexed Metalens for Compact Computational Spectral Imaging," Laser & Photonics Reviews **18**, (2024).
38. H. Xia, B. Chen, C. Zhang, C. Zhu, C. Wang, and Z. Zheng, "Joint optimization of coded aperture metasurface and residual self-attention network for snapshot full-Stokes imaging," Opt. Express **32**, 29609 (2024).
39. W. Du, X. Liu, Z. Wang, Z.-L. Deng, T. Ma, X. He, J. Qi, Z. Hu, X. Liu, and Q. Yang, "PSF-engineered snapshot full-Stokes polarizing-spectral-imaging via metasurface with reciprocally encoded anisotropic detour phase," Optics & Laser Technology **181**, 111645 (2025).
40. J. Wen, S. Feng, Y. Chen, W. Shi, H. Gao, Y. Shao, Y. Shao, Y. Zhang, W. Shen, and C. Yang, "Full-Stokes Spectro-Polarimetric Camera with Full Spatial Resolution," Laser & Photonics Reviews **19**, (2025).
41. X. Dun, H. Ikoma, G. Wetzstein, Z. Wang, X. Cheng, and Y. Peng, "Learned rotationally symmetric diffractive achromat for full-spectrum computational imaging," Optica **7**, 913 (2020).
42. Y. Cai, J. Lin, Z. Lin, H. Wang, Y. Zhang, H. Pfister, R. Timofte, and L. V. Gool, "MST++: Multi-stage Spectral-wise Transformer for Efficient Spectral Reconstruction," in *2022 IEEE/CVF Conference on Computer Vision and Pattern Recognition Workshops (CVPRW)* (IEEE, 2022), pp. 744–754.
43. "U-Net: Convolutional Networks for Biomedical Image Segmentation," in *Lecture Notes in Computer Science* (Springer International Publishing, 2015), pp. 234–241.
44. K. He, X. Zhang, S. Ren, and J. Sun, "Deep Residual Learning for Image Recognition," in *2016 IEEE Conference on Computer Vision and Pattern Recognition (CVPR)* (IEEE, 2016), pp. 770–778.
45. F. Yasuma, T. Mitsunaga, D. Iso, and S. K. Nayar, "Generalized Assorted Pixel Camera: Postcapture Control of Resolution, Dynamic Range, and Spectrum," IEEE Transactions on Image Processing **19**, 2241–2253 (2010).
46. B. Arad and O. Ben-Shahar, "Sparse Recovery of Hyperspectral Signal from Natural RGB Images," in *Computer Vision – ECCV 2016*, B. Leibe, J. Matas, N. Sebe, and M. Welling, eds., Lecture Notes in Computer Science (Springer International Publishing, 2016), Vol. 9911, pp. 19–34.
47. Y. Li, Q. Fu, and W. Heidrich, "Multispectral illumination estimation using deep unrolling network," in *2021 IEEE/CVF International Conference on Computer Vision (ICCV)* (IEEE, 2021), pp. 2652–2661.
48. X. Hu, Y. Cai, J. Lin, H. Wang, X. Yuan, Y. Zhang, R. Timofte, and L. Van Gool, "HDNet: High-resolution Dual-domain Learning for Spectral Compressive Imaging," in *2022 IEEE/CVF Conference on Computer Vision and Pattern Recognition (CVPR)* (IEEE, 2022), pp. 17521–17530.
49. F. A. Kruse, A. B. Lefkoff, J. W. Boardman, K. B. Heidebrecht, A. T. Shapiro, P. J. Barloon, and A. F. H. Goetz, "The spectral image processing system (SIPS)—interactive visualization and analysis of imaging spectrometer data," Remote Sensing of Environment **44**, 145–163 (1993).